# A Review of the SolarWinds Attack on Orion Platform using Persistent Threat Agents and Techniques for gaining Unauthorized Access

Antigoni Kruti
Northumbria University
London, UK
Antigoni.kruti@northumbria.ac.uk

Usman Butt
Engineering and Environment
Northumbria University
London, UK
Usman.Butt@northumbria.ac.uk

Rejwan Bin Sulaiman
Engineering and Environment
Northumbria University
London, UK
Rewjan.sulaiman@northumbria.ac.uk

**Abstract –** This paper of work examines the SolarWinds attack, designed on Orion Platform security incident. It analyses the persistent threats agents and potential technical attack techniques to gain unauthorized access. In 2020 SolarWinds attack indicates an initial breach disclosure on Orion Platform software by malware distribution on IT and government organizations such as Homeland Security, Microsoft and Intel associated with supply chains leaks consequences from small loopholes in security systems. Hackers increased the number of infected company's' and businesses networks during the supply-chain attack, hackers were capable to propagate the attack by using a VMware exploit. On the special way they started to target command injections, privilege escalations, and use after free platforms of VMware. In this way, they gained access to Virtual Machines and in the east way pivot other servers. This literature review aim to analyse the security gap regarding to SolarWinds incident on Orion Platform, the impact on industry and financial sectors involving the elements of incident response plan. Therefore, this research paper ensures specifications of proper solutions for possible defence security systems by analysing a SolarWinds attack case study via system evaluation and monitoring. It concludes with necessary remediation actions on cyber hygiene countermeasures, common vulnerabilities and exposure analysis and solutions.

## 1. INTRODUCTION

### 1.1 Background

The United States of America acknowledged by the end of 2020s a sophisticated cyberattack with large impact in the SolarWinds company. This company provides software examination infrastructure tools and network computer monitoring tools all over the world [1]. The company has the "Orion" system, that possesses advantaged access of data and log activities with wide implementation of privileges. In this way, SolarWinds is lucrative target by hackers, that globally affected the 300.000 major organizations in the sectors of military, telecom, and governance.

Cyber security experts rely on VM sandbox environment to develop admin privileges connected with company's network. VMware offers products and solutions to manage virtual machines and applications.

VMware illegal activities have been accomplished in several assaults, involving the SolarWinds attack. In 2020 the VMware's site reported 30 vulnerabilities, one of the significantly used was CVE-2020-4006 (Common Vulnerabilities and Exposure) for privileges escalation on the breach of SolarWinds. According to Huddleston in 2021 The potential attack outcome was the compromising of 18,000 SolarWinds costumers by allowing the cloud services access as Microsoft Office 365 and government agencies [2].

The SolarWinds attack is considered to be complex, devastating and well-crafted supply chain cyber-attack [1]. It compromised the dynamic file of companies and the virus was disseminated by unknown vendor to customers networks. Threat actors consists of resolving sub-domains of C2 domain avsvmcloud.com

by Domain Generation Algorithms (DGAs) while, preventing to induce suspicions on outbound traffic. The temporary file replacement approach involves a utility to executed the payload of employees [3].

## 1.2 Motivation and contribution

This paper of work contributes on the reported solutions analysis of the malware to emphasise the development of innovative technology tools such as Configuration Change Management and Vulnerability Assessment CIP-010-3 standard and CIP-004-6 to detect and disable unauthorize access of cyber systems and to minimize the risk of system compromising by plans implementations related with Personnel and trainings documented processes.

## 1.3 Paper organization

This paper of research will critically analyse the anatomy of the SolarWinds attack in Orion network management software centralized on IT monitoring authority and access on clients' system. It will treat the impacts and remediation actions of the specific vulnerabilities. A section of detailed methodologies will be dedicated on a proper section, elaborated with proper security responses with cyber-hygiene to mitigate the potential damage. The second section consist of the SolarWinds attack anatomy and performance tools. The thirds sections involves analysis of used techniques to develop the attack such as command injection and use-after-free. The forth section develops cyber hygiene solutions and future solutions tools and techniques such as System Security Management (SSM) CIP-007-6 standard, Secure Location Protocol (SLP), Configuration Change Management and Vulnerability Assessment CIP-010-3 standard. The fifth section is the last section that concludes with countermeasures to optimize and to prevent the leak of SolarWinds attack.

# 2. LITERATURE REVIEW

## 2.1 The Anatomy of SolarWinds attack

This section provides a deep analysis of the SolarWinds supply chain attack focused on the examination of cybers kill chain stages, which is supported by figure 1 illustrations. The first step is Reconnaissance, it consists of data collection from all email database of employees [2].The attacker has aimed to implement techniques to exploit an authentication vulnerability on mail server and to launch the attack by developing a profile for the targeted victims.

The second phase is the Weaponization, the attackers created the Dynamic Link Library (DLL) backdoor executed on the "SolarWinds.BusinessLayerHost.exe" program [4]. It is known as Solorigate for Microsoft company and Sunburst for FireEye company.

The next stage is the delivery, the hackers implemented the technique of Spear phishing to take advantages form Orion network management software company. They disguised the malware through an updated company version, which could be distributed on the supply chain.

Furthermore, on the fourth stage of the attack the hackers waited for the exploitation of the updated network management software of Orion with the malware (SolarWinds.Orion.Core.BusinessLayer.dll with a hash file of [b91ce2fa41029f6955bff20079468448] and 2. C:\ WINDOWS \ SysWOW64 \ netsetupsvc.dll – This file that is not noticed on the Windows operating system [5].

Moreover, Installation is the fifth stage of the SolarWinds attack, the victims could install the managing software updated and the malware open a backdoor on server victim to automatically install the malware itself for hackers to get access on server services.

The last stage is Command and Control. In a typical infection of the Windows domain name, the first 14 characters would be sent to the Command-and-Control server by Sunburst [6]. Then, the hackers instructed Sunburst to deliver statuses of the security product sending the information to enable Sunburst to launch more secondary HTTP-based robust.

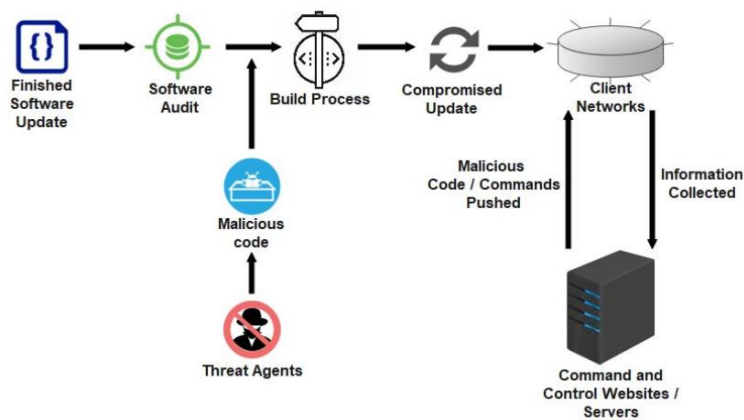

**Figure 1: The supply chain attack illustration**

## 2.2 The SolarWinds performance outcome on costumers

The offensive of the SolarWinds attack began in 2019 and the figure 2 depicts main tools and techniques, that are used from hackers to successfully achieve the SolarWinds attack. There are 18,000 users affected by the SolarWinds supply chain attack exposed by the risk of crucial sensitive and private data leaking of employee [7].

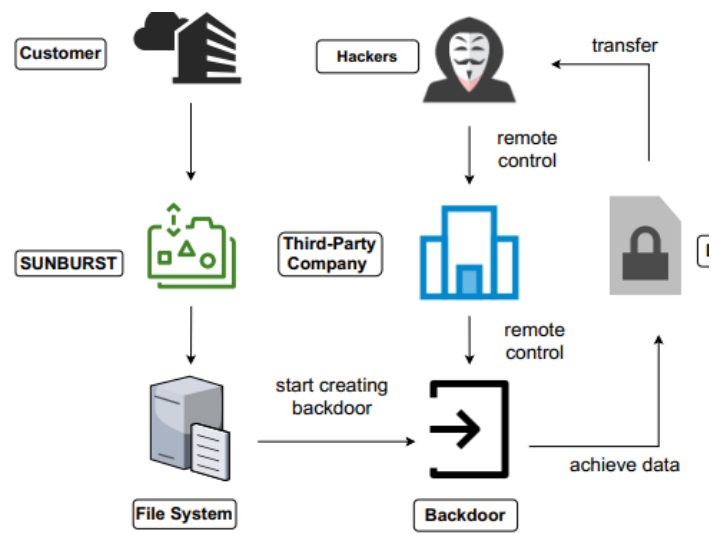

**Figure 21: The trojanized tools and techniques usage to send information**

FireEye analysis outcome emphasis the masquerades of the malware network traffic called the Orion Improvement Program (OIP) protocols and collect investigational outcomes blended with SolarWinds activity into plugin configuration files. Furthermore, the blocklists are stopped to recognize any antivirus tools to run into services, processes, and drivers by the trojanized plug-in platform known as SolarWinds.Orion.Core.BusinessLayer.dll [2]. During the execution of trojanized platform plug-ins attackers minimize their footprints and create remote access with credential usage lateral movement on the targeted systems. Attackers utilized Teardrop as a malware dropper to lead directly to the memory avoiding leaving behind any trace on the disk from attackers modified utilities on targeted systems [8]. In this part of the attack the VMware exploits allow hackers inside the compromised devices to plant web shells to laterally move and steal sensitive data from Microsoft ADFS servers. According to United States Government Accountability Office in 2022 hackers apply CVE-2020-4006 to get access to Workspace ONE software data hosted to generate Security Assertion Markup Language (SAML) credentials by taking advantages from Microsoft encryption algorithm vulnerabilities [4]. It allows hackers to extract sensitive data, certifications and source code of Microsoft programs and give footholds to attack other vulnerable systems.

# 3. THE SOLARWINDS ATTACK TECHNIQUES AND VULNERABILITIES

## 3.1 Command injection

Hackers implemented command injection techniques on SolarWinds attack because of CVE-2020-4006 weakness. It granted the privileges escalation, that affected six products of VMware related to VMware Workspace One Access Connector, Identity Manger Connector, VMware Identity Manager, Suite Lifecycle Manager and VMware Cloud Foundation [9]. The Common Vulnerability Score System Calculator as 9.1 before the attack and then downgraded to 7.2 score, it was recognized that the administrative configurator of outside resource would not be changed. Hence, the computer that is using the compromised affected system could not be harmed by the weakness, but the VMs are exploited to be hacked and changed because of administrative configuration management, the network compromised server is the starting point of the attack [10]. Attackers could exploit the running port of VMware configurator system and to have credentials valid access on admin current account permissions to run their code as a gateway to connect web shells on VMs. Attackers could use packet sniffing, keyloggers social engineering methods and phishing [11]. The case of VMware's of CVE-2020-4006 includes specific functionalities of disabilities for the configurator operations on port 8443 [4].

## 3.2 Use-after-free

Use after free is a vulnerability in VMware ESXi versions 6.5, 6.7 and 2.0 was CVE-2020-3992. It could be exploited by hackers accessing port 427 on network access machine [4]. User after free vulnerabilities is focused on the allowed amount of code substitution. Attackers replaced the dynamic memory allocated by a pointer with their own commands and code. This vulnerability allows hackers to realize the encryption of several VMs' hard drives.

CVE-2020-4004 is another uses after free vulnerability within XHCI USB controller. There are two vulnerabilities CVE-2020-4004 and CVE-2020-4005,that mixed together allow a malicious actor to access VMX process privileges of escalation to exploit the chain. Furthermore, those vulnerabilities could enable code execution on the host device as it could be the VMX process [3]. The VMX's system call management flaw could escalate because of CVE-2020-4005 vulnerability.

# 4. THE PROPOSED SOLUTIONS AND COUNTERMEASURES

## 4.1 Remediation action on Cyber hygiene countermeasures

This section highlights the generic countermeasures by following the Federal Entity North American Electric Reliability Corp (NERC) to prevent the SolarWinds attack [12]. The NERC Critical Infrastructure Protection (CIP) standards provide reliability, sustainability, and security systems from four important standards to the SolarWinds attacks such as CIP-010-3, CIP-004-6, CIP-005-6, CIP007-6 [6]. Configuration Change Management and Vulnerability Assessment CIP-010-3 standard detects and disable unauthorize access to cyber systems. On the other hand, CIP-004-6 standard minimizes the risk of system compromising by plans implementations related with Personnel and trainings documented processes [10]. It is important to apply non-reused passwords to counteract brute-forcing of credentials, for instance, VMware could be controlled by including setting proper permissions for authorized access and last privilege maintaining [13].

The entities apply the given instructions of CIP depend on the plan purposes. The Electronic Security Perimeter (ESP) CIP-005-6 defines security parameters to decrease the level of instability on cyber systems [11]. The primary reasons of SolarWinds assault was the IT network authentication and the network development were not disconnected, it was one way authentication for both networks [14]. If SolarWinds applied the CIP-00506 instructions to separate network development from IT network authentication, the attack might be avoided. [12]

Another standard is System Security Management (SSM) CIP-007-6 standard, that is used to manage the security system by technical, procedural, and operational requirements specifications [15]. Furthermore, the Encapsulation Security Payload (ESP) protocol traces the control risks overs services and takes responsibilities for data packets encryption across a Virtual Private Network (VPN), it minimizes the threats from attacks and increase the security level for development network [9].

## 4.2 CVE-2020 solutions

| Attack Vector | Description | Possible Defense |
|---|---|---|
| CVE-2020-4006 | OS Command Injection | Update to patched version, or implement workaround |
| CVE-2020-3992 | Use-after-free | Update to patched version, or disable SLP |
| CVE-2020-4004/CVE-2020-4005 | Use-after-free/privilege escalation | Update to patched version, or Remove XHCI USB Controller |

**Table 1: CVE-2020 description of solutions**

CVE-2020-4006 could be mitigated in two ways such as the update of patched version and implementation of workaround supplied from VMware. In this way the affected service could be disable [16]. CVE-2020-3992 could be solved by affected software updating and Secure Location Protocol (SLP) disable on the impacted server. Hence, CVE-2020-4004 and CVE-2020-4005 could be solved by updating the malicious server and removing USB controller processes by port disable because of vulnerabilities requirement of physical assessment to USB port [17]. On this way, the use-after-free vulnerabilities and the escalation of privileges will be decreased [18].

## 4.3 Future proposed solutions

The Security Operation Centre 1 (SOC) compliance helps organizations to develop and analyse audit reports of information assurance regarding security, integrity, confidentiality, and privacy activities [19]. It is supported by American Institute of Certified Public Accounts (AICPA) and ISAE 3402 standard or SSAE 18 standard help in SOC 1 development [20]. On the other hand, SOC 2 ensures risk vendor management and internal governance of the company by covering a timeframe of 6-12 months for all technological and operational maintenance providers based on industry-specific penetration testing procedures compliance [21].

# 5. CONCLUSIONS

This paper of work examined the SolarWinds attack tools, methods and techniques with the help of virtual machine vulnerabilities. It illustrated the main goals to ensure thee convenient decision making to optimise those vulnerabilities such as privileges escalation or use-after-free to be exploited on VMs. The combination between several traffic anomalies detections and flow packets tagging tools and techniques are crucial to detect and to mitigate on real-time the software supply chain attacks [22].

The Orion Platform attack has shown the methods of threat manipulations with multiple techniques to destabilize the targeted machine system by effective infiltrations tools [31]. However, there is still an opportunity to improve the detection measures and to decrease the risk of SolarWinds assault with the help of specific technical analysis supported by Cyber Hygiene maintenance and CVE solutions [24]. This will avoid hackers to gain linking computer software for malicious activities commitment on several organizations such as FireEye, CISCO and Microsoft.